\documentclass[
]{ceurart}

\usepackage{tabularx}
\usepackage{multirow}
\usepackage{listings}

\usepackage[frozencache,cachedir=.]{minted}
\setminted{frame=lines,
    framesep=2mm,
    fontsize=\footnotesize,
    breaklines=true,
    linenos
    }
\usepackage{listings}
\begin{document}

\copyrightyear{2024}
\copyrightclause{Copyright for this paper by its authors.
  Use permitted under Creative Commons License Attribution 4.0
  International (CC BY 4.0).}

\conference{32nd Irish Conference on Artificial Intelligence and Cognitive Science, December 9-10, Dublin, Ireland}

\title{AICat: An AI Cataloguing Approach to Support the EU AI Act}



\author[1]{Delaram Golpayegani}[%
orcid=0000-0002-1208-186X,
email=sgolpays@tcd.ie,
]

\address[1]{ADAPT Centre, Trinity College Dublin, Dublin, Ireland}

\author[2]{Harshvardhan J. Pandit}[%
orcid=0000-0002-5068-3714,
email=me@harshp.com,
]

\address[2]{ADAPT Centre, Dublin City University, Dublin, Ireland}

\author[1]{Dave Lewis}[%
orcid=0000-0002-3503-4644,
email=delewis@tcd.ie,
]


\begin{abstract}
  The European Union's Artificial Intelligence Act (AI Act) requires providers and deployers of high-risk AI applications to register their systems into the EU database, wherein the information should be represented and maintained in an easily-navigable and machine-readable manner. Given the uptake of open data and Semantic Web-based approaches for other EU repositories, in particular the use of the Data Catalogue vocabulary Application Profile (DCAT-AP), a similar solution for managing the EU database of high-risk AI systems is needed. This paper introduces \textbf{AICat}—an extension of DCAT for representing catalogues of AI systems that provides consistency, machine-readability, searchability, and interoperability in managing open metadata regarding AI systems. This open approach to cataloguing ensures transparency, traceability, and accountability in AI application markets beyond the immediate needs of high-risk AI compliance in the EU. AICat is available online at \url{https://w3id.org/aicat} under the CC-BY-4.0 license.
\end{abstract}


\maketitle

\section{Introduction}

The European Union (EU) Artificial Intelligence Act (AI Act)~\cite{aiact}, which entered into force on 1 August 2024, stands as a landmark legal regime for development and use of AI. 
Within the AI Act, there is a high demand for Regulatory Technology (RegTech) solutions to serve the foundational and technical backbone required for implementation and enforcement of the Act~\cite{fehlinger2023enabling}. Drawing parallels with the EU digital regulations, notably the General Data Protection Regulation (GDPR)~\cite{gdpr}, and looking into the body of compliance and enforcement solutions proposed in regard to such regulations suggest adoption of Semantic Web for effective and scalable compliance and enforcement solutions.

In the context of the AI Act, one area where the Semantic Web is anticipated to be used is the implementation of the \emph{EU database of high-risk AI systems}. The database, which is to be established and managed by the European Commission in collaboration with Member States, is intended to encompass information regarding high-risk AI systems as declared by their providers and deployers.
From the technical perspective, the AI Act requires the information contained within the database to be \emph{``easily navigable''} and \emph{``machine-readable''} (Art. 71 (4)), with different levels of accessibility, i.e. publicly and non-publicly accessible.

To implement and manage the EU database, and any catalogues of AI-related resources, a layer of metadata is needed to facilitate cross-referencing, traceability, transparency, interoperability, and comparability. The current state of existing repositories of AI systems, models, and datasets shows that adoption of machine-readable metadata is limited (see  \autoref{relatedwork}). In this paper, we address this gap in AI repositories as well as the pressing need for the EU to establish the high-risk AI database by proposing \textbf{AICat} as a cataloguing approach. AICat extends the Data Catalog Vocabulary (DCAT)~\cite{dcat-v3}, 
 enabling describing AI systems and components, including AI models and datasets, in catalogues through a consistent, standardised, and interoperable mechanism. This leads to the contributions of this work as: 
 \begin{enumerate}
     \item An in-depth analysis of the AI Act's registration requirements for providers and deployers of high-risk AI systems,
     \item \textbf{AICat}, an extension of DCAT that provides a mechanism for cataloguing AI systems and their incorporating components in registries of AI systems, including but not limited to the EU database of high-risk AI systems.
 \end{enumerate}

\section{Related Work} \label{relatedwork}

With the proliferation of AI models, systems, and use cases, open AI repositories and commercial marketplaces have been created to facilitate the discovery and sharing of resources~\cite{kumar2021marketplace}. This section investigates the literature to identify related studies that address registering and sharing metadata about AI systems and their risks, in particular within the EU database of high-risk AI systems, using Semantic Web technologies.  

Currently, there are a few well-known repositories that leverage metadata for describing resources. The \textbf{Hugging Face Hub}\footnote{\url{https://huggingface.co/docs/hub/index}} is a centralised repository of open-source models and datasets, wherein each model or dataset is accompanied with metadata describing them. This enables discovery, sharing, and filtering of resources available on Hugging Face's Model and Data Hubs through the use of open JSON-based metadata. 
The Hub contains a repository of Spaces, i.e. ML demo apps, which, unlike the Model and Data Hub, do not support the inclusion of documented information and structured metadata. 
Similarly, \textbf{Kaggle} provides repositories of  Datasets\footnote{\url{https://www.kaggle.com/datasets}} and Models\footnote{\url{https://www.kaggle.com/models}}, where datasets, models, and generative AI applications are indexed and documented using detailed Data and Model Cards. Using the Kaggle repository, datasets and models can be published, shared, tagged, searched, and sorted. Compared to Hugging Face Data Hub which supports indexing only open-source resources, Kaggle Datasets allows for sharing metadata about both proprietary and publicly available datasets. 
The \textbf{AI-on-Demand (AIoD)  platform}\footnote{\url{https://aiod.eu/}} is a European-funded project that serves as a community-driven hub for cataloguing AI-related solutions and components that contribute to the European ecosystem of AI excellence and trust. AIoD's asset catalogue\footnote{\url{https://www.ai4europe.eu/research/ai-catalog}} covers a wide range of resources including datasets, libraries, ML models, AI services, tools, use cases, and even tutorials. AIoD also provides JSON-based metadata for describing resources\footnote{\url{https://api.aiod.eu/redoc}}.  

\textbf{Croissant} \cite{croissant2024}  is a framework developed by MLCommons—a non-profit open AI engineering consortium that enables expressing metadata for datasets with a focus on information that is essential in machine learning workflows. The Croissant vocabulary\footnote{\url{https://docs.mlcommons.org/croissant/docs/croissant-spec.html}} is an extension of \texttt{schema.org/Dataset} vocabulary for metadata of ML datasets, which is expressed in the JSON-LD format. The Croissant framework is supported by a user-friendly tool to assist non-technical users in creation and modification of metadata. Although it is not a dataset repository, it has been integrated with existing data repositories, including HuggingFace, adding a layer of metadata.

 While the information in the aforementioned registries is mostly presented in semi-structured formats such as JSON, none of them follow standardised approaches for data sharing or cataloguing. In regard to standardised approaches, the Data Catalog Vocabulary (DCAT)~\cite{dcat-v3}—the W3C's recommended vocabulary for publishing data catalogues—and particularly its application profile for data portals in Europe (DCAT-AP)~\cite{dcat-ap} have been adopted by the European Commission to promote open, standardised, and interoperable data sharing, prominently in the European Data Portal (EDP)\footnote{\url{https://data.europa.eu/en}}, which is the central point of access to open data provided by the EU's public agencies \cite{fabin2019dcatap}. 
  Recently, \textbf{MLDCAT-AP}~\cite{mldcat-ap} has been proposed as an extension of DCAT-AP for including information about machine learning models in data catalogues. One of the distinguishing features of MLDCAT-AP is inclusion of information about \emph{risks} associated with ML models. 
  
  Of relevance to the contributions of this work is the \textbf{Data Processing Catalogue (DPCat)}~\cite{ryan2022dpcat}, which is an extension of DCAT and DCAT-AP that enables representing, maintaining, and exchanging ROPA\footnote{Register of Processing Activities}-related information in the form of datasets and catalogues. DPCat further enables creating documentation to address the GDPR's ROPA requirements.

\autoref{tab:<ch2-repository-comparison>} shows a comparison of existing approaches for cataloguing AI, models, and datasets. Currently, providing metadata, typically in JSON format, regarding datasets and models is an established practice. However, there is little attention to cataloguing AI systems and consequently there is no standardised machine-readable vocabulary that supports cataloguing of AI systems as well as their incorporating components.

\begin{table}[!h]
    \centering
    \caption{Comparison of AI cataloguing approaches (a black circle (\textbullet) indicates the criterion is satisfied, while a blank circle ($\circ$) indicates that it is not)}
    \begin{tabularx}{\textwidth}{|p{5.5cm}|p{4cm}|X|X|}
    \hline
        \textbf{Work} & \textbf{Scope} & \textbf{Format of metadata} &\textbf{ Use of standardised vocabularies}  \\
    \hline 
    \rowcolor[HTML]{EFEFEF} \multicolumn{4}{|l|}{\textbf{Repositories}}\\
    \hline
   
          \textbf{Hugging Face Data/Model Hub} & Dataset/Model &  JSON & $\circ$ \\

     \rowcolor[HTML]{EFEFEF}
           \textbf{Kaggle dataset/model repository} & Dataset/Model & HTML & $\circ$ \\

          \textbf{AI-on-Demand (AIoD) platform} & AI assets (dataset, model, services) & JSON & $\circ$ \\
\hline
         \rowcolor[HTML]{EFEFEF}
    \multicolumn{4}{|l|}{\textbf{Approaches}}\\
\hline   
           
         \textbf{Croissant} \cite{croissant2024} & Dataset & JSON-LD & $\circ$ \\
    
    \rowcolor[HTML]{EFEFEF}    
         \textbf{MLDCAT-AP}~\cite{mldcat-ap} & ML models & JSON-LD & \textbullet \\
         \textbf{DPCat}~\cite{ryan2022dpcat} & GDPR's ROPA & Turtle & \textbullet \\

     \hline
    \end{tabularx}
    \label{tab:<ch2-repository-comparison>}
\end{table}

\section{Analysis of the AI Act's Registration Requirements} \label{aiact_analysis}

Under the AI Act, providers and deployers of Annex III high-risk AI systems and providers of non-high-risk Annex III systems, i.e. systems that meet the conditions of Annex III but are considered as non-high-risk by the provider, are required to register their systems into the EU database (Article 49). According to Article 71, the EU database should be set up and maintained by the European Commission, in collaboration with the Member States. It shall be \emph{``accessible and publicly available''} (with some exceptions), provided in a \emph{``user friendly manner''}, and should be \emph{``easily navigable and machine-readable''}. The EU database aims to act as an instrument for the Commission and the Member States to facilitate monitoring the current uptake of Annex III AI systems—regardless of their associated risk category—within the EU market and to serve as a transparency measure for sharing information regarding such systems with the public (Article 71 and Recital 131). The EU database therefore is a key data interoperability point between the Commission, AI providers, AI deployers, and the public.

\autoref{tab:<ch3-aiact-registration>} provides a summary of the registration provisions specified in Article 49. As shown in Table, the list of information elements that should be registered and their level of openness, i.e. publicly accessible or not, depends on the role of the registrant and the type of the system. In this, notably, submitting information regarding incorporating AI models, whether they are general-purpose or not, is not needed. However, information about general-purpose AI models should be made available to downstream AI providers that intend to use the model within their systems (Article 53). 

Annex VIII, wherein the information to be submitted upon the registration of high-risk AI systems is outlined, was analysed to identify the \emph{general} information that should be provided when registering an AI system into the EU database. Detailed information, such as the system's logic, instructions for use, and summary of fundamental rights impact assessment are not included, due to their descriptive nature and the lack of guidelines. In addition, for the general description of the general-purpose AI model, the key information elements listed in Annex XII, Point 1, were included to enable representation of AI components. \autoref{tab:<information-req>} shows the key information elements extracted from Annex VIII and XII.

\begin{table}[!h] 
    \centering
    \small
    \caption{Registration requirements for high-risk AI systems under the EU AI Act}
    
    \begin{tabularx}{\textwidth}{|p{1cm}|p{3cm}|X|X|X|p{3cm}|}
    \hline
       
        \textbf{AI Act Article} & \textbf{AI System} & \textbf{Where?} & \textbf{What Information?} & \textbf{Who?} & \textbf{When?} \\
        \hline

      \rowcolor[HTML]{EFEFEF}
        49(1) & High-risk as per Annex III, P. 3, 4, 5, 8 & Public EU database & Annex VIII (A)  & AI provider or authorised representative & Before placing on the market or putting into service \\
       
        49(1) \& (4) &  High-risk as per Annex III, P. 1, 6, and 7 & Non-public EU database & Annex VIII (A), points 1 to 10 (except 6, 8, and 9) & AI provider or authorised representative & Before placing on the market or putting into service  \\
       \rowcolor[HTML]{EFEFEF}
        49(2) & Meets Annex III, P. 2, 3, 4, 5, 8 conditions but non-high-risk as per assessment of the provider & Public EU database & Annex VIII (B) & AI provider or authorised representative & Before placing on the market or putting into service \\
     
        49(2) \& (4) & Meets Annex III, P. 1, 6, \& 7  conditions but non-high-risk as per assessment of the provider & Non-public EU database & Annex VIII (B), points 1 to 5 \& points 8 \& 9 & AI provider or authorised representative & Before placing on the market or putting into service \\
     \rowcolor[HTML]{EFEFEF}
        49(3) & High-risk as per Annex III, P. 3, 4, 5, 8 & Public EU database & Annex VIII (C) & AI deployer (public authorities, Union institutions, bodies, offices, or agencies) & Before putting into service or using \\
    
        49(3) \& (4) & High-risk as per Annex III, P. 1, 6, \& 7 & Non-public EU database & Annex VIII (C), points 1 to 3 & AI deployer (public authorities, Union institutions, bodies, offices, or agencies) & Before putting into service or using \\
      \rowcolor[HTML]{EFEFEF}
        49(5) & High-risk as per Annex III, P. 2 & Register at national level & Not mentioned & Not mentioned & Not mentioned \\
    \hline
    \end{tabularx}
   \label{tab:<ch3-aiact-registration>}

\end{table}


\begin{table}[!h]
    \centering
    \caption{Key information elements to be registered into the EU database}
    \begin{tabularx}{0.8\textwidth}{|p{1.5cm}|p{2cm}|X|}
    \hline
       \textbf{Annex}&\textbf{Clause} & \textbf{Requirement}  \\
       \hline
        \rowcolor[HTML]{EFEFEF} \multicolumn{3}{|l|}{Information about \textbf{operators}, including providers and deployers} \\
       \hline
        & A1, B1 & AI provider's name \\
        & A1, B1 & AI provider's address\\
        & A1, B1 & AI provider's contact details  \\
        VIII& C1 & AI deployer's name \\
        & C1 & AI deployer's address  \\
       & C1 & AI deployer's contact details  \\
        \hline
         \rowcolor[HTML]{EFEFEF}
        \multicolumn{3}{|l|}{Information about \textbf{AI system}} \\
        \hline
        & A4, B4 &
        AI system's trade name  \\
        & A4, B4 & AI system's additional reference \\
        & A5, B5 & AI system's intended purpose \\

        VIII & A7, B8 & AI system's market status   \\
        & A10, B9 & Countries where system is available  \\
        
        \hline
         \rowcolor[HTML]{EFEFEF}
        \multicolumn{3}{|l|}{Information about \textbf{components}, i.e. datasets and models}\\
        \hline
        & A6 & Data used by the system \\
        VIII & A6 & Input data used by the system \\
        & A5, B5 & Component's intended purpose \\
        & --& AI models used within the system \\
        \hline
        & 1-1b & Model's use policy \\
     XII   & 1-1c & Model's date of release \\
   
     & 1-1g & Model's input data  \\
   
      & 1-1g & Model's output data  \\   
  
        & 1-1h & Model's license \\
        \hline

    \end{tabularx}
    \label{tab:<information-req>}
\end{table}

\newpage
\section{AICat - a DCAT Extension for Cataloguing AI Systems} \label{section:<ch5-aicat>}

AICat is an application profile for specifying catalogues of AI systems that avails a thin layer of metadata to enhance interoperability and cross-referencing within the AI value chain. Building upon DCAT version 3, which supports cataloguing varying resources, AICat enables providing metadata about AI systems, models, and datasets. In addition to the resources that are already used by DCAT, AICat reuses existing concepts from our previous work, including the Data Privacy Vocabulary (DPV)\footnote{\url{https://w3id.org/dpv/}}~\cite{pandit2024dpv}, its technology extension\footnote{\url{https://w3id.org/dpv/tech}}, and the AI Risk Ontology (AIRO)\footnote{\url{https://w3id.org/airo}}~\cite{golpayegani2022airo}.

The key information elements identified from the AI Act's registration obligations, discussed in \autoref{aiact_analysis}, shape the functional requirements of AICat. These requirements, which are expressed in the form of competency questions following the methodology described in \cite{figueroa2009orsd}, are shown in \autoref{tab:<ch5-aicat-orsd>}.

\begin{table}[!h]
\caption{AICat profile requirements specification}
\centering
\scriptsize
\begin{tabular}{|c|c|c|}
\hline
\multicolumn{3}{|c|}{\cellcolor[HTML]{c7c7c7}\textbf{AICat Requirements Specification Document}} \\ \hline

\multicolumn{3}{|c|}{\cellcolor[HTML]{EFEFEF}\textbf{1. Purpose}} \\ 
\hline
\multicolumn{3}{|p{0.95\textwidth}|}{The purpose of the AICat profile is to use DCAT and AIRO to describe catalogues of AI systems and their associated components, such as datasets and AI models.} \\ 
\hline

\multicolumn{3}{|c|}{\cellcolor[HTML]{EFEFEF}\textbf{2. Scope}} \\ \hline
\multicolumn{3}{|p{0.95\textwidth}|}{The scope of AICat is limited to the \emph{atomic} information that should be submitted upon the registration of high-risk AI systems into the EU database, outlined in Annex VIII. This means that descriptive information, for instance the system's logic and findings of the fundamental rights impact assessment, is not included in the scope.}  \\ 
\hline
\multicolumn{3}{|c|}{\cellcolor[HTML]{EFEFEF}\textbf{3. Implementation Language}} \\ 
\hline
\multicolumn{3}{|p{0.95\textwidth}|}{OWL, DCAT} \\
\hline
\multicolumn{3}{|c|}{\cellcolor[HTML]{EFEFEF}\textbf{4. Key Uses}} \\
\hline
\multicolumn{3}{|p{0.95\textwidth}|}{
USE 1. Maintaining and managing metadata about AI systems, datasets, and models in interoperable and standardised catalogues. \newline
USE 2. Discovering and comparing AI solutions. \newline
USE 3. Cataloguing and sharing information about AI systems with the public in a transparent manner. This includes the use by the European Commission for sharing metadata of the high-risk AI systems indexed in the EU database.} 
\\
\hline

\multicolumn{3}{|c|}{\cellcolor[HTML]{EFEFEF}\textbf{5. Ontology Requirements}}\\ 
\hline
\multicolumn{3}{|c|}{\cellcolor[HTML]{EFEFEF}\textbf{a. Non-Functional  Requirements}} \\
\hline
\multicolumn{3}{|p{0.95\textwidth}|}{NFR 1. AICat shall be published online with standard documentation.  \newline
NFR 2. AICat shall reuse concepts and relations from existing ontologies, including AIRO, to the fullest extent possible. }
\\ \hline
\multicolumn{3}{|c|}{\cellcolor[HTML]{EFEFEF}\textbf{b. Functional  Requirements: Groups of Competency Questions}}\\ \hline
\multicolumn{1}{|p{0.30\textwidth}|}{\cellcolor[HTML]{EFEFEF}\textbf{CQG1. AI systems}} &
\multicolumn{1}{|p{0.30\textwidth}|}{\cellcolor[HTML]{EFEFEF}\textbf{CQG2. Datasets}} &
\multicolumn{1}{|p{0.30\textwidth}|}{\cellcolor[HTML]{EFEFEF}\textbf{CQG3. AI models}} \\
\hline
\multicolumn{1}{|p{0.30\textwidth}|}{ 
CQ1-1. What is the name of the system?  \newline
CQ1-2. Who is the system's provider? \newline
CQ1-3. Who is the system's deployer? \newline
CQ1-4. What is the system's intended purpose? \newline
CQ1-5. What is the system's market availability status? \newline
CQ1-6. In which countries is the system made available? \newline
CQ1-7. What are the additional references to the system? \newline}
&
\multicolumn{1}{|p{0.30\textwidth}|}
{
CQ2-1. Which datasets are used by the system?\newline
CQ2-2. What is the system's input data?\newline
CQ2-3. What is the dataset's use policy?
} 

&

\multicolumn{1}{|p{0.30\textwidth}|}
{CQ3-1. Which models are used by the system? \newline
 CQ3-2. What is the model's release data? \newline
 CQ3-3 What is the model's input data?
 \newline
 CQ3-4 What is the model's output data?\newline
 CQ3-5. What is the model's license? \newline
 CQ3-6. What is the model's use policy?
 } \\
\hline

\end{tabular}

\label{tab:<ch5-aicat-orsd>}
\vspace{-0.1in}
\end{table}

\subsection{AICat Overview}
AICat extends DCAT version 3~\cite{dcat-v3}, since this version of DCAT supports cataloguing resources beyond datasets. By extending DCAT, AICat aims to scale the cataloguing to include AI systems and models to address the needs of the EU database. 
\autoref{tab:<aicat-spec>} illustrates how the identified requirements are mapped into concepts from DCAT, AIRO, DPV, and DPV's TECH extension (for prefixes refer to \autoref{lst:<ch5-aicat-example>}). As shown in the Table, the \emph{intended purpose} of a system is represented as a policy modelled using the AI Use Policy (AIUP) profile~\cite{golpayegani2024aiup}, which is an extension of the Open Digital Rights Language (ODRL)~\cite{odrl}, for expressing AI use offers, requests, and agreements between multiple parties across the AI value chain. 

\bigskip
\begin{table}[!h]
    \centering
    \footnotesize
    \caption{Specifications for representing AI systems and models in AICat}
    \begin{tabularx}{\textwidth}{|p{0.53cm}|p{1.4cm}|X|p{4.2cm}|p{2.7cm}|}
    \hline
    \textbf{CQ}& \textbf{AI Act Annex}& \textbf{Requirement} & \textbf{Metadata Field} & \textbf{Range} \\
       \hline
       
        \multicolumn{5}{|l|}{\cellcolor[HTML]{EFEFEF}Information about \textbf{AI system}}\\
        \hline
        1-1&  VIII, A4 \& B4 & AI system's trade name & \texttt{dct:title} & \texttt{rdfs:Literal}  \\
        \hline
        1-2& VIII, A1 \& B1 & Provider's information & \texttt{airo:isProvidedBy}  &  \texttt{airo:AIProvider}  \\
        \hline
        1-3 & VIII, C1 & Deployer's information  & \texttt{airo:isDeployedBy} & \texttt{airo:AIDeployer}  \\
         \hline
        1-4& VIII, A5 \& B5 & AI system's intended purpose & \texttt{odrl:hasPolicy} & \texttt{aiup:UsePolicy} \\
        \hline
       1-5 &  VIII, A7 \& B8 & AI system's market status  & \texttt{tech:hasMarketAvailab- ilityStatus} & \texttt{tech:MarketAva- ilabilityStatus}  \\
         \hline
      1-6&   VIII, A10 \& B9 & Countries where system is available & \texttt{dpv:hasCountry} & \texttt{dpv:Country}  \\
        \hline
     1-7 &   VIII, A4 \& B4 & AI system's additional reference  & \texttt{dct:isReferencedBy} &  \texttt{dcat:Resource}\\
    
        \hline
        \multicolumn{5}{|l|}{\cellcolor[HTML]{EFEFEF}Information about \textbf{components}}\\
        \hline
      2-3 &  VIII, A5 \& B5&   Component's intended purpose  & \texttt{odrl:hasPolicy} & \texttt{aiup:UsePolicy} \\
        \hline
        \multicolumn{5}{|l|}{\cellcolor[HTML]{EFEFEF}Information about \textbf{datasets}}\\
        \hline
       2-1& VIII, A6 & Data used by the  system or model
       &\texttt{airo:hasTrainingData}, \texttt{airo:hasValidationData}, \texttt{airo:hasTestingData} & \texttt{airo:Data}  \\
         \hline
       2-2&  VIII, A6 & Input data used by the system & \texttt{airo:hasInput}& \texttt{airo:Data} \\
         \hline
         \multicolumn{5}{|l|}{\cellcolor[HTML]{EFEFEF}Information about \textbf{models}}\\
         \hline
       3-1&  --& AI models used within the system & \texttt{airo:hasModel} & \texttt{airo:AIModel}\\
        \hline
       
     3-2&  XII, 1-1c &Model's date of release & \texttt{dct:issued} & \texttt{xsd:date}  \\
      \hline
      3-3 &  XII, 1-1g & Model's input data & \texttt{airo:hasInput}& \texttt{airo:Data} \\
      \hline
      3-4 & XII, 1-1g & Model's output data & \texttt{airo:hasOutput} & \texttt{airo:Data} \\
      \hline
      3-5& XII, 1-1h &  Model's license & \texttt{airo:hasLicense} & \texttt{airo:License} \\
      \hline
       3-6 &  XII, 1-1b & Model's use policy & \texttt{odrl:hasPolicy} & \texttt{aiup:UsePolicy} \\
        
        \hline

    \end{tabularx}
    \label{tab:<aicat-spec>}
\end{table}
\bigskip

\begin{figure}[h!]
    \centering
    \fbox{\includegraphics[width=\linewidth]{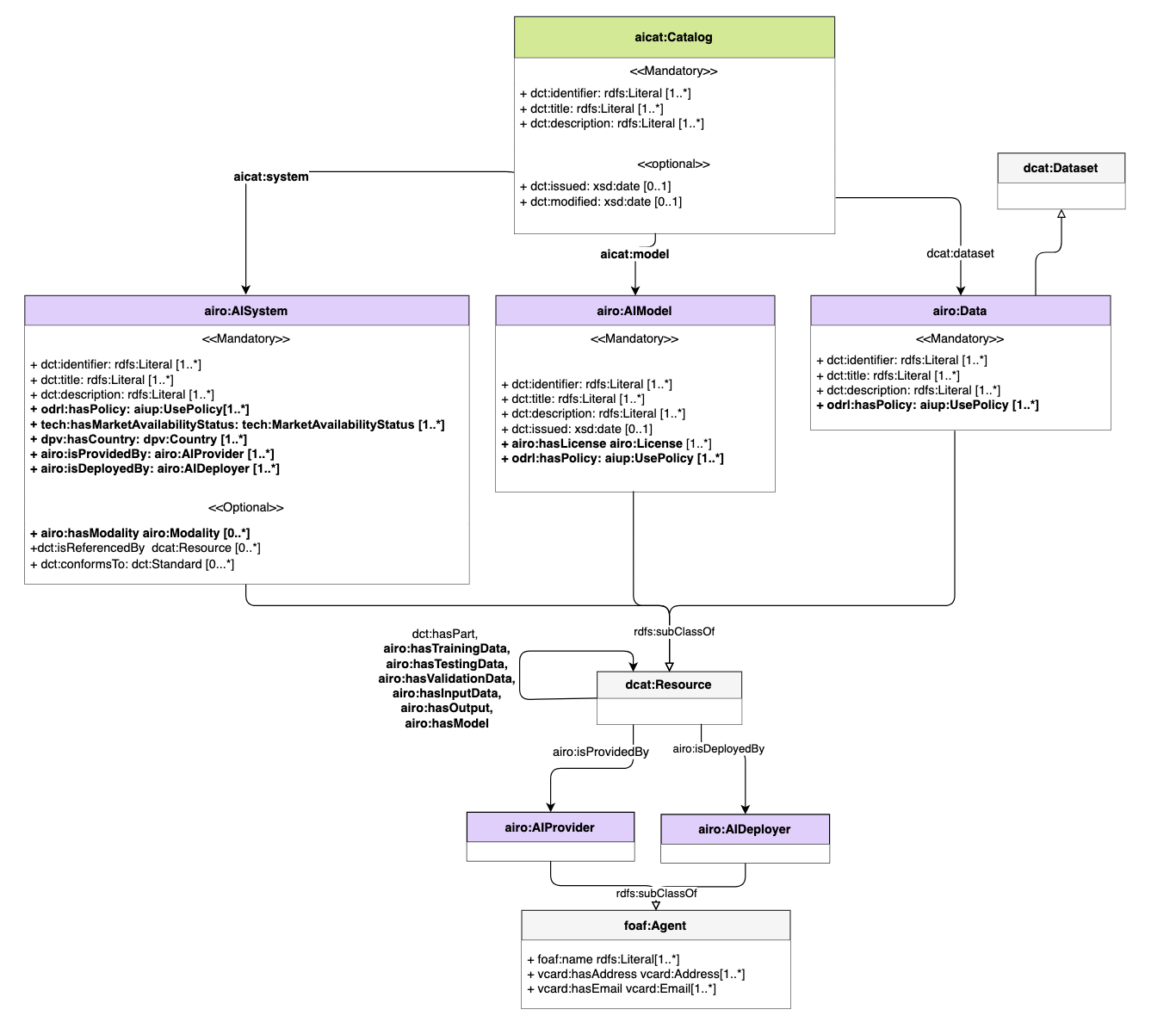}}
    \caption{An overview of the AICat Profile}
    \label{fig:<ch5-aicat>}
\end{figure}

\autoref{fig:<ch5-aicat>} depicts an overview of AICat's information model. As illustrated in the Figure, \texttt{aicat:Catalog} is a sub-class of  \texttt{dcat:Catalog} that provides a curated collection of metadata about AI systems, models, and datasets. AICat extends DCAT by introducing \texttt{airo:AISystem} and \texttt{airo:AIModel} as sub-classes of \texttt{dcat:Resource}, enabling inclusion of their metadata in an \texttt{aicat:Catalog}. Given that \texttt{airo:Data} is a sub-class of \texttt{dcat:Dataset}, cataloguing data is also supported by AICat. While the inclusion of AI systems was directly linked to the scope of the EU database, whose aim is to index AI systems, the inclusion of models and datasets was driven by the existing focus in the state of the art on cataloguing these AI components, as reviewed in \autoref{relatedwork}. 

\texttt{aicat:system}, \texttt{aicat:model}, and \texttt{dcat:dataset} are sub-properties of \texttt{dcat:resource} that allow linking the catalogue to the resources indexed therein. 
To enable modelling the relationships between the resources, for example to show which datasets used for training a model, \texttt{airo:hasTrainingData}, \texttt{airo:hasTestingData}, \texttt{airo:hasValidationData}, \texttt{airo:hasInput}, \texttt{airo:hasOutput}, and \texttt{airo:hasModel} are reused from AIRO. 

AICat's documentation was generated using WIDOCO~\cite{garijo2017widoco} and is available online at \url{https://w3id.org/aicat} under the CC-BY-4.0 license.

By following DCAT-AP~\cite{dcat-ap}, AICat can further distinguish between \emph{mandatory}, \emph{recommended}, \emph{optional}, and \emph{deprecated} elements based on the requirements of the AI Act. Even though implementing such normative profiles can easily be realised by defining the aforementioned property types for each of the information elements, in the context of the AI Act, identification of whether provision of an information element is mandatory, recommended, optional, or deprecated requires additional guidelines and codes of conduct.

\newpage
AICat is introduced as a minimal extension of DCAT. This extension introduces the \texttt{aicat:Catalog} class and its relations with \texttt{airo:AISystem} and \texttt{airo:AIModel}, both added as new types of \texttt{dcat:Resource}.
  One of the key directions for improving AICat is using the Shapes Constraint Language (SHACL)~\cite{shacl} to specify the level of necessity for information elements—which can be mandatory, recommended, or optional. \autoref{lst:<ch5-aicat-shacl>} shows an example of a SHACL shape indicating that each AI system should have at least one provider. Currently, AICat does not define such a normative profile due to the absence of recommendations and guidelines in regard to the AI Act.

\begin{listing}[H]
    
\begin{minted}{turtle}
@prefix sh: <http://www.w3.org/ns/shacl#> .
@prefix airo: <https://w3id.org/airo#> .
:AIProviderShape a sh:NodeShape;
                                       sh:targetClass airo:AISystem ;
                                       sh:property [
                                         a sh:PropertyShape ;
                                         sh:path airo:isProvidedBy;
                                         sh:minCount 1 ] .
\end{minted}
\caption{Example of a SHACL shape that specifies the requirement for presence of at least one provider for an AI system}
\label{lst:<ch5-aicat-shacl>}
\end{listing}

\newpage
\subsection{Proof-of-Concept Implementation} 

To illustrate an example of cataloguing, we use an example of an AI-based student proctoring system, described in  \cite{panigutti2023role, hupont2023usecasecards, golpayegani2024AIcards}. The system, which is called \emph{Proctify}, is provided by \emph{AIEduX} and intended to detect suspicious behaviour during online exams by analysing facial behaviour. The system incorporates a facial analysis toolkit, provided by a third party, to analyse a person's facial information, including the head pose, gaze direction, and face landmarks' positions. This extracted information is then provided as an input to a model, \emph{SusBehavedModel}, which has been trained in-house by the system's provider using \emph{SusBehavedDataset}, to determine whether the student is displaying suspicious behaviour.
\autoref{lst:<ch5-aicat-example>} presents a summarised version of an \texttt{aicat:Catalog} that contains metadata about \emph{Proctify} and its components. 
As shown in the Listing, the policies for using the AI system and its incorporating components are expressed using the AIUP profile.

\begin{listing}

\begin{minted}{turtle}
@prefix xsd: <http://www.w3.org/2001/XMLSchema#> .
@prefix rdfs: <https://www.w3.org/TR/rdf12-schema/> .
@prefix rdf: <http://www.w3.org/1999/02/22-rdf-syntax-ns> .
@prefix dct: <http://purl.org/dc/terms/> . 
@prefix dcat: <https://www.w3.org/TR/vocab-dcat-3/> .
@prefix dpv: <https://w3id.org/dpv#> .
@prefix tech: <https://w3id.org/dpv/tech#> .
@prefix airo: <https://w3id.org/airo#> .
@prefix aiup: <https://w3id.org/aiup#> .
@prefix aicat: <https://w3id.org/aicat#> .
@prefix ex: <http://example.com/proctify#> .

ex:aieduX-catalogue-01 a aicat:Catalog, dcat:Catalog  ;
    dct:identifier "aiedux-cat01"^^xsd:string ;
    dct:title "AIEduX catalogue"@en ;
    dct:description "AI systems and models provided by AIEduX"@en ;
    dct:created "2024-05-05"^^xsd:date ;
    dcat:dataset ex:susbehaved_dataset ; 
    aicat:model ex:susbehaved_model;
    aicat:system  ex:proctify . 

ex:susbehaved_dataset a dcat:Dataset, airo:Data ;
   dct:identifier "aiedux-d012"^^xsd:string ;
   dct:title "SusBehavedDataSet"@en ;
   dct:description ".. includes suspicious behaviour data.."@en ;
   odrl:hasPolicy ex:susbehaved_dataset_policy  .

ex:susbehaved_model a dcat:Resource, airo:AIModel ;
    dct:identifier "aiedux-m022"^^xsd:string ;
    dct:title "SusBehavedModel"@en ;
    dct:description ".. determines suspicious behaviour .."@en ;
    dct:issued "2024-02-15"^^xsd:date ; 
    airo:hasTrainingData ex:susbehaved_dataset ;
    odrl:hasPolicy ex:susbehavedmodel_policy . 
    
:proctify a dcat:Resource, airo:AISystem ;
    airo:isProvidedBy ex:aiedux ;
    dct:identifier "aiedux-ai031"^^xsd:string ; 
    dct:title "Proctify"@en ;
    dct:description "An AI-based proctoring system..."@en ;
    tech:hasMarketAvailabilityStatus tech:MarketAvailable ;
    dpv:hasCountry <http://dbpedia.org/resource/Italy> ;
    dcat:contactPoint <http://example.org/aieduX-AI031/contact> ; 
    airo:hasModel ex:susbehaved_model ;
    odrl:hasPolicy ex:proctify_use_policy .   

 ex:susbehaved_dataset_policy a aiup:UseOffer  .
 ex:susbehavedmodel_policy a aiup:UseOffer .
 ex:proctify_use_policy a aiup:UseOffer .
\end{minted}
\caption{An example of \texttt{aicat:Catalog} for describing a catalogue}
\label{lst:<ch5-aicat-example>}
\end{listing}

\newpage
\section{Potential Benefits of AICat}
In terms of potential benefits, through reusing widely-used W3C standardised vocabularies, the AICat enables expressing metadata regarding AI systems and AI components within catalogues, wherein common vocabularies and open linked data-based formats are used. 
Therefore, the AICat addresses the AI market needs for a consistent and interoperable mechanism for cataloguing AI solutions~\cite{aiwatch2022publicsectorreport}, in a way that enables federated search and comparison across AI, model, and data catalogues offered by different vendors—a crucial feature often required in AI procurement processes. In relation to this, the European Commission's dataset of selected uses of AI in the public sector~\cite{jrc-use-of-ai-dataset} is a prominent resource, whose interoperability and searchability can be enhanced through adoption of a cataloguing mechanism such as AICat.  

At the organisational level, AICat could assist AI providers and deployers in providing structured catalogues of AI systems and components. At the European level, a similar approach to AICat is expected to be adopted for the implementation of the database of high-risk AI systems as required by Article 71 of the AI Act. Given that AICat ensures traceability while protecting privacy by providing metadata without revealing sensitive information within a database, it supports the implementation of the non-public section of the EU database and provides a structure for registration forms. AICat potentially addresses the gap in the European open data portal in providing FAIR (Findable, Accessible, Interoperable, and Reusable) information regarding existing AI systems and models provided or deployed by public organisations. 
AICat also has the potential to promote cross-border interoperability required by the recently-enforced Interoperable Europe Act~\cite{eu-interoperabilityact}, particularly in the implementation of the \emph{Interoperable Europe portal}—the EU's single point of entry for information related to cross-border interoperability of trans-European digital public services (Interoperable Europe Act, Article 8). In this, AICat can be employed to facilitate sharing information and best practices to support interoperability in public procurement of AI-based solutions.

Compared to existing cataloguing approaches, reviewed in \autoref{relatedwork}, AICat expands the scope of cataloguing to AI systems. From this literature review, MLDCAT-AP~\cite{mldcat-ap} bears a close resemblance to AICat, especially in the use of DCAT. MLDCAT-AP has been supported by the European Commission's Semantic Interoperability Community (SEMIC), and therefore it might be a candidate to be adopted in the implementation of the EU database. However, prior to this, it needs to be extended to include specifications of AI systems in the catalogue in alignment with the requirements of the AI Act. This can be realised by the integration of MLDCAT-AP and AICat. Another key feature of MLDCAT-AP, in comparison with AICat, is the inclusion of risk information in the catalogue. While AICat can support DCAT-based documentation of risks by reusing \texttt{airo:hasRisk}, in its current form it does not go beyond the general, non-descriptive information elements of Annex VIII, mainly due to the absence of related official guidelines. 

\section{Conclusion and Further Work}

In this paper, we proposed AICat as a novel technical solution for cataloguing AI systems in an open, machine-readable, and interoperable format based on the evolving requirements of the AI value chain, particularly the requirements of the EU AI Act. Using AICat facilitates discovery, integration, and sharing information associated with AI systems and components amongst the stakeholders involved in the AI value chain based on the existing proven mechanism of (open) data portals. 

By demonstrating this solution, we hope that similar open and interoperable approaches will be adopted in the implementation of the AI Act, in particular the creation of the EU database of high-risk AI systems as per Article 71. Our work also contributes to trustworthy and responsible use of AI by enabling creation of scalable and interoperable AI catalogues on the internet by using a unified and coherent vocabulary.

\begin{acknowledgments}
This project has received funding from the European Union’s Horizon 2020 research and innovation programme under the Marie Skłodowska-Curie grant agreement No 813497 (PROTECT ITN). The ADAPT SFI Centre for Digital Media Technology is funded by Science Foundation Ireland through the SFI Research Centres Programme and is co-funded under the European Regional Development Fund (ERDF) through Grant\#13/RC/2106\_P2. 
\end{acknowledgments}

\bibliography{aicat_ref}

\appendix

\end{document}